# Core-to-Core X-ray Emission Spectra from Wannier Based Multiplet Ligand Field Theory

Charles A. Cardot, Joshua J. Kas, Jared E. Abramson, John J. Rehr, and Gerald T. Seidler

Physics Department, University of Washington, Seattle WA

Abstract

Recent advances using Density Functional Theory (DFT) to augment Multiplet Ligand Field Theory (MLFT) have led to *ab-initio* calculations of many formerly empirical parameters. This development makes MLFT more predictive instead of interpretive, thus improving its value for understanding highly correlated $3d$, $4d$, and $f$-electron systems. Here, we explore a DFT + MLFT based approach for core-to-core Kα x-ray emission spectra (XES) and evaluate its performance for a range of transition metal systems. We find good agreement between theory and experiment, as well as the ability to capture key spectral trends related to spin and oxidation state. We also discuss limitations of the model in the context of the remaining free parameters and suggest directions forward.

## I. Introduction

The qualitative connection between ground state electronic structure and macroscopic physical properties of molecules and condensed phases has been evident since the earliest work based on the Fermi-Dirac distribution [1], Bloch waves [2], and the quantum mechanical treatment of chemical bonding [3]. However, excited-state electronic structure is necessary for a quantitative connection between theory and many experimental probes. This is especially apparent for core shell spectroscopies, such as electron energy loss spectroscopy (EELS) [4], x-ray photoelectron spectroscopy (XPS) [5], and x-ray absorption spectroscopy (XAS) (including x-ray absorption fine structure, XAFS), nonresonant x-ray emission spectroscopy (XES), and resonant inelastic x-ray scattering (RIXS) [6]. Therefore, the accurate simulation of these probes has long been a core goal of the theoretical condensed matter and chemistry communities [7, 8]. The ability to reliably predict these core-shell spectroscopies has clear scientific benefit for interpretation of experiments on a case-by-case basis. Also, accurate and efficient approaches for solution of the 'forward problem' of predicting spectra from structure has served as the foundation for more recent machine (ML) approaches [9-12], leading to new analyses and also to new directions for the high-throughput science being enabled by improved synchrotron facilities.

Many theoretical and computational approaches for core shell spectroscopy have been developed [13], including those based on time-dependent density functional theory (TDDFT), [14], many-body perturbation theory in the form of the Bethe-Salpeter equation [15, 16], multiple scattering methods [17], quantum chemistry based methods [5, 18, 19], and the multiplet ligand field theory (MLFT) approach which we adopt here [7, 20-22]. With comparatively lower computation cost, the application of MLFT to highly correlated materials finds good agreement with a variety of x-ray spectroscopies, e.g., XAS [23], XES [24, 25], XPS [26], and RIXS [27, 28]. Although the many free parameters used in fitting MLFT to experiment limits its predictive value, recent developments utilizing Wannier functions have allowed for *ab-initio* DFT based predictions of most of the free parameters related to solid state effects [20]. This process is laid out in Figure 1, where the modified Slater-Condon parameters, tight-binding Hamiltonian ($H^{TB}$), and some charge-transfer terms are extracted from DFT calculations, greatly simplifying the adjustable phase space of the model, allowing for a much more predictive approach to MLFT [29-32].

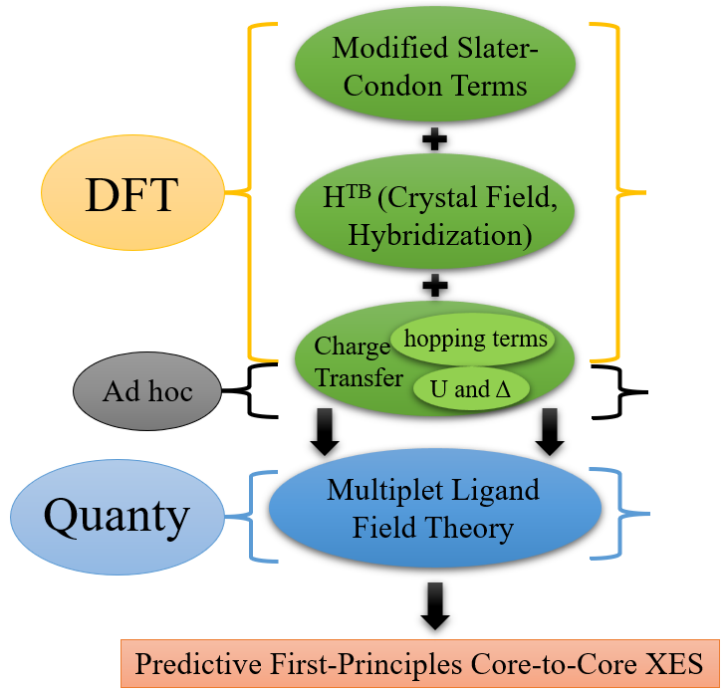

**Figure 1**: Schematic representation of the theory workflow: DFT is used to extract *ab-initio* values for many traditionally empirical parameters, leaving only a few terms which must be set *ad-hoc*. These are all combined within the MLFT framework, as implemented with Quanty and then used to calculate the XES of the target system.

Our goal in this work is to explore the DFT + MLFT approach, which focuses on the parameters that are most directly connected to the perturbed atomic picture, making it quite appropriate for core-to-core XES. Although DFT + MLFT been successfully used to simulate XAS, XPS, and RIXS, its application to core-to-core XES has, as far as we know, been unexplored. We present a first test of DFT + MLFT for Kα ($2p \rightarrow 1s$) XES for a variety of $3d$ transition metal (TM) compounds. The choice of emission line and chemical systems is appropriate for a validation of the approach in correlated systems. First, MLFT is fundamentally a perturbation theory, and as such, we aim to validate the leading order of environmental perturbation. This condition is justified by the weak coupling between the $2p$ states and the ligand-level electronic structure. The locality of Kα XES has previously been contrasted to the more extended nature of XAFS, with the observation that Kα spectra can have generally simpler sensitivity to atomically-derived observables such as oxidation state [33]. Second, the collection of $3d$ TM systems studied includes some materials with strongly localized valence-level electrons where MLFT is natural, and some systems with strong covalency where empirical parameters in MLFT typically take unexpected, if not unphysical, values to obtain numerical agreement with experiment. Again, the point here is to challenge the leading-order calculation of deep core-to-core XES via DFT + MLFT.

This paper is organized as follows: In section II, we summarize the principal theory of nonresonant-XES and the details of the DFT + MLFT framework as well as methods used to reduce the number of free parameters down to only two. The details of the codes being used are documented elsewhere [34, 35], but example input files are provided in SI-I. In section III, we present results, discuss the validity of the approximations, and compare calculated spectra with experimental spectra for a range of transition metal compounds. We examine performance across spin-state, environmental symmetry, and oxidation. Finally, in section IV, we explore the remaining two-dimensional phase space, discuss systematic drawbacks, and propose future directions for this framework.

## II. Theoretical Formalism

### A. XES Theory

The creation of a non-resonant diagram line in x-ray fluorescence is illustrated in Figure 2. In core-to-core XES, a core electron absorbs a high energy photon, leaving a deep core hole behind and ejecting a photoelectron. After a few fs, the core hole is filled by a less tightly bound (shallower) electron, and a photon is emitted, leaving the system in a final state with either a semi-core or valence level hole. When studied with modest energy resolution, the resulting x-ray fluorescence is commonly used for elemental identification [36]. On the other hand, following the synchrotron community convention, the same radiation is called x-ray emission when it is studied with energy resolution comparable to intrinsic broadening. Such XES experiments provide information about the element specific chemical and electronic environment, often including sensitivity to oxidation and spin state of the species of interest [6].

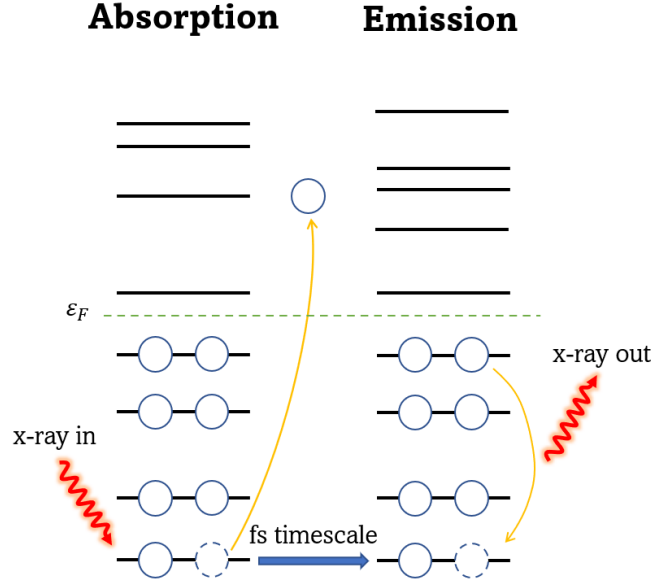

**Figure 2:** A diagrammatic depiction of the non-resonant x-ray fluorescence process. Going from left to right, an x-ray is absorbed by a core electron, which is subsequently ejected from the system into the continuum. A short time later a less tightly bound electron decays into the hole left behind by the deep core electron, emitting an x-ray to preserve energy.

Formally, the complete photon-in, photon-out process can be described by the Kramers-Heisenberg formula [7, 37],

$$\frac{d\sigma(\omega_1,\omega_2)}{d\omega_1 d\omega_2} \propto \sum_F \left|\sum_M \frac{\langle F|\hat{T}_2|M\rangle\langle M|\hat{T}_1|I\rangle}{E_I + \omega_1 - E_M + i\Gamma_M/2}\right|^2 \frac{(\Gamma_F/2\pi)}{(E_I - E_F + \omega_1 - \omega_2)^2 + \Gamma_F^2/4}, \quad (1)$$

where $|I\rangle, |M\rangle, |F\rangle$ are initial, intermediate, and final N-electron many-body states, with corresponding energies $E_I, E_M, E_F$. The broadenings $\Gamma_M$, $\Gamma_F$, are due to the lifetimes of the intermediate and final states, respectively, and $\hat{T}_1, \hat{T}_2$, are dipole transition operators. The terms $\omega_1$ and $\omega_2$ are the energies of the incoming and outgoing photons, respectively, making $\omega_1 - \omega_2$ the energy transferred to the system. For nonresonant XES, the spectral shape is independent of the incoming photon energy. In this case, the photoelectron can be neglected and the spectrum is proportional to the integral over incoming energy $\omega_1$ [7]. If the lifetime broadening $\Gamma_m$ is assumed constant, the integration simplifies to,

$$\frac{d\sigma_{XES}(\omega_2)}{d\omega_2} \propto \sum_f \left| \sum_m \frac{\langle f|\hat{T}_2|m\rangle\langle m|\hat{s}|I\rangle}{E_f - E_m - \omega_2 + i(\Gamma_m + \Gamma_f)/2} \right|^2 (\Gamma_m + \Gamma_f)/2\pi. \quad (2)$$

The states $|m\rangle, |f\rangle$ are N - 1 electron many-body states, i.e., they do not include the photoelectron, and the $\hat{T}_1$ dipole transition operator has been replaced with the 1s annihilation operator $\hat{s}$. No polarization term is included in $\hat{T}_1$ because the continuum final state of the photoelectron is inconsequential. It should be noted that while we are focusing on the case of Kα XES for the sake of being explicit, the approach described here is generalizable to any core-to-core XES. A detailed derivation for going from Eq. 1 to Eq. 2 is given in chapter 8 of de Groot and Kotani [7]. For Kα XES, the lifetime broadenings $\Gamma_m$ and $\Gamma_f$ are approximately constant and are equal to the core hole broadenings of the 1s and 2p shells ($\Gamma_s$ and $\Gamma_p$), and $\hat{T}_2 = \sum_{s,p}(\vec{\epsilon}\cdot\vec{r})\hat{s}^\dagger\hat{p}$ is limited to the dipole transition operator connecting the six 2p fermionic modes to the two 1s fermionic modes. Eq. 2 considers all intermediate states which connect the final and initial states and is referred to here as *the two-step approach*.

Eq. 2 is commonly used when discussing nonresonant XES [25]; however, another possible approximation is to take the XES equivalent of the final-state rule used in valence-to-core (VTC) XES [38] and XAS [7, 30]. In this case, the sum over intermediate states in Eq. 2 is assumed to be dominated by the ground state in the presence of the deep-core hole $|i'\rangle$ [39], which gives,

$$\frac{d\sigma_{XES}^{K\alpha}(\omega_2)}{d\omega_2} \propto \sum_f \left| \frac{\langle f|\hat{T}_2|i'\rangle}{E_f - E_{i'} - \omega_2 + i(\Gamma_s + \Gamma_p)/2} \right|^2 (\Gamma_s + \Gamma_p)/2\pi. \quad (3)$$

This approximation simplifies the problem considerably, as it requires only a single sum over the final states, and it will henceforth be referred to as the *one-step approach*. This can also be reformulated using a one-body Green's function formalism as shown in supplemental information section SI-II. A full comparison between the two approaches will be explored in section IV.A.

## B. Multiplet Ligand Field Theory

For the calculations here we use an MLFT Hamiltonian which is the sum of an atomic contribution and a tight binding Hamiltonian, $H = H^{atom} + H^{TB}$. The atomic Hamiltonian accounts for the deep-core ($1s$, $2p$) states as well as many-body Coulomb interactions $U_{ijkl}$ between core and $d$ states, and between individual $d$ states. The tight-binding Hamiltonian $H^{TB}$ describes the single particle energies of the $d$ and ligand states, as well as hopping between them. These two terms are given by

$$H^{atom} = \sum_i \epsilon_i \hat{n}_i + \sum_{d_1 d_2 d_3 d_4} U_{d_1 d_2 d_3 d_4} \hat{c}^\dagger_{d_1} \hat{c}^\dagger_{d_2} \hat{c}_{d_3} \hat{c}_{d_4} + \sum_{d_1 d_2 p_1 p_2} U_{d_1 p_1 d_2 p_2} \hat{c}^\dagger_{d_1} \hat{c}^\dagger_{p_1} \hat{c}_{p_2} \hat{c}_{p_2}$$
$$+ \sum_{d_1 d_2 s_1 s_2} U_{d_1 s_1 d_2 s_2} \hat{c}^\dagger_{d_1} \hat{c}^\dagger_{s_1} \hat{c}_{s_2} \hat{c}_{p_2} + \sum_{p_1 p_2} h^{SO}_{p_1 p_2} \hat{c}^\dagger_{p_1} \hat{c}_{p_2} + \sum_{d_1 d_2} h^{SO}_{d_1 d_2} \hat{c}^\dagger_{d_1} \hat{c}_{d_2}, \quad (4)$$

$$H^{TB} = \sum_{\gamma \gamma'} V^{LF}_{\gamma \gamma'} \hat{c}^\dagger_\gamma \hat{c}_{\gamma'}. \quad (5)$$

In these equations, the index $i$ indicates a sum over all states, $s$ and $p$ a sum over the $1s$ or $2p$ states of the absorbing metal atom, $d$ a sum over absorber $3d$ states, and $\gamma$ a sum over the $3d$ and ligand states. $\epsilon_i$ is the non-relativistic single particle energy of orbital $i$, $U_{ijkl}$ are the Coulomb matrix elements, and $V^{LF}_{\gamma \gamma'}$ (taken from the DFT tight binding Hamiltonian) encompasses the crystal field as well as couplings between the $d$ and ligand states. Finally, $h^{SO}$ is the single-particle spin orbit coupling Hamiltonian $h^{SO} = \vec{l} \cdot \vec{s}$. The $U_{ijkl}$ Coulomb terms are parameterized by the Slater-Condon $F$ and $G$ parameters corresponding to direct and exchange Coulomb interactions, as well as the average screened Coulomb interactions $U_{dd}, U_{pd}$, and $U_{sd}$ which encompass the spherically symmetric contribution to the multipole expansion [7]. Finally, the charge transfer energy $\Delta$ defines the energy required to transfer one electron from the ligand to the metal. The Slater-Condon parameters are also obtained from a separate DFT calculation, so that the only remaining free parameters are the charge-transfer energy $\Delta$ and the average screened Coulomb interactions $U_{dd}, U_{pd}$, and $U_{sd}$. More details on the form and definition of these parameters can be found in [40], [41], and in the supplemental information section SI-II.

## III. Methods

### A. Computational

In this work we use the full-potential local-orbital electronic structure code FPLO [35] to calculate the tight binding Hamiltonian and the radial wavefunctions necessary for the Slater-Condon *F* and *G* terms. The code Quanty [34] was used for the subsequent solution of the MLFT Hamiltonian and for the construction of the spectrum [20]. The interface between FPLO and Quanty was built by Heinze and Haverkort [40] following the framework developed by Haverkort et al. [20]. We used FPLO version 14.00-49-x96_64, with the Perdew-Wang 92 exchange correlation functional. Unless otherwise stated, default convergence parameters were used for the density and total energy, $10^{-6}$ Å$^{-3}$ and $10^{-8}$ Hartree respectively, and convergence with respect to the *k*-mesh was carefully checked.

Using band structure as a guide, the energy window of the down-projection of the Wannier functions was tailored to the TM-3*d* and ligand-2*p* orbitals. It should be emphasized that the Wannier functions calculated within FPLO are not maximally localized. The fact that we use a tight binding Hamiltonian then requires that, along with a localization procedure that reduces the number of overall terms in the Hamiltonian, the Wannier functions are chosen such that they reproduce both the band energies and the orbital character of the bands. The details of this step are explained in reference [42]. The DFT mean-field Coulomb interaction between the 3*d* electrons is subtracted from the tight-binding Hamiltonian to prevent double counting; this is accounted for later with the $U_{dd}$, $U_{pd}$, $U_{sd}$, and $\Delta$ free parameters. Only 2*p*-3*d* exchange Slater-Condon parameters are considered in the Kα XES calculation, neglecting the weak 1*s*-3*d* exchange terms. Additionally, it is well known that $U_{pd}$ and $U_{sd}$ are close numerically, and are usually 0.5 eV – 2.0 eV larger than $U_{dd}$ [41]. Thus, we use the relations $U_{pd}/U_{dd} = 1.15$ and $U_{sd}/U_{dd} = 1.20$. This leaves only 2 free parameters: $U_{dd}$ and $\Delta$, making it reasonable to do a full exploration of the remaining adjustable parameter space for each system.

It is important to compare theory with known trends in Kα XES. As an example we test the commonly asserted linear relationship between the Kα$_1$ full width half-maximum (FWHM) and the number of unpaired 3*d* electrons [43]. Mulliken population analysis is often used for determining the 3*d* unpaired occupation [44], but this can yield unphysical results for diffuse systems. Instead, we calculate the expectation value of the $S^2$ operator acting on the intermediate ground state wavefunction, which is consistent with the established MLFT framework. From this it is trivial to solve the equation,

$$S(S+1) = \langle \hat{S}^2 \rangle; \quad n_{unpaired\ 3d} = 2S, \qquad (6)$$

and hence extract the number of unpaired 3d electrons for the system.

## B. Experimental

MnO, $Mn_2O_3$, $MnO_2$, and $CrCl_2$ samples were prepared from greater than 97% reagent grade stock (Sigma Aldrich). All Mn samples were prepared by pressing a 1:1 mass mixture of powder sample and hexagonal boron nitride (BN) into a 13-mm diameter pellet before being encased in a polyimide pouch. The air and moisture sensitive $CrCl_2$ sample was prepared by sealing powder in a quartz tube (0.01 mm wall thickness), in a nitrogen glovebox.

The XES spectra for these samples were collected using the laboratory-based spectrometer described in detail in Jahrman, et al., [45]. Briefly, using a conventional x-ray tube (Varex VF80, Pd-Anode) operated at 100 W electron beam power (35 kV, 2 mA) samples were illuminated for 1.5 hours and 14 hours for the Mn samples and $CrCl_2$ samples respectively, insuring at least 10,000 counts at the $K\alpha_1$ peak for all spectra. Additional $Cr_2O_3$ and $PbCrO_4$ experimental spectra are from Jahrman, et al., [46] and the NiO spectra is digitized from Kawai, et al. [47].

## C. Data Analysis

To compare spectral features quantitatively between experiment and theory and across samples, we have extracted the $K\alpha_1$ FWHMs, $K\alpha_2$ FWHMs, and $K\alpha_1$:$K\alpha_2$ integral ratios. $K\alpha$ XES is complicated by the presence of shoulders and additional multiplet features within $K\alpha_1$ and $K\alpha_2$ peaks, usually making it necessary to use more than two peak functions (Lorentzian or Voigt) to achieve the most accurate fit to a spectrum [48]. Therefore, to extract the $K\alpha$-FWHM parameters, we have used linear interpolation between data points which lie on either side of the half-maximum crossing points following Lafuerza et al., [24]. For the $K\alpha_1$:$K\alpha_2$ integral intensity ratio, individual peaks must be extracted from the overall spectra to accurately measure their relative areas. This is done using two to four Voigt functions to achieve an approximate line shape, from which we extract integral intensities. The fitting procedure follows Jabua [49] and Voigt fits for all compounds are presented in the supplemental information section SI-V.

Unless otherwise stated, all spectra have been integral normalized and shifted to align the maximum of $K\alpha_1$ at zero. Also, all calculated spectra are broadened using lifetime values in Campbell and Papp [50], which are reported for each main peak in Table 1. A 1.0 eV Gaussian broadening for experimental resolution was applied to all spectra except that of MnF$_2$, where 0.5 eV was used due to the higher resolution of the spectrometer used in that experiment [48]. Additional information about the extraction of the FWHM and how the spectra were broadened can be found in supplemental information section SI-V.

## IV. Results and Discussion

This section is organized as follows. In IV.A we address the validity of the one-step approach compared to the two-step approach and demonstrate that they produce nearly identical results. All subsequent calculations use the one-step approach which greatly reduces computational costs. In IV.B we compare theory and experiment for complete spectra and address the limitations of our approach by system-specific discussions. In IV.C we compare theory and experiment for a few spectral parametrizations, including the FWHM of the two peaks, and the K$\alpha_1$:K$\alpha_2$ intensity ratio. We also show how a key trend involving the K$\alpha_1$ FWHM and number of unpaired 3$d$ electrons is well reproduced by theory. Finally, in IV.D we address the remaining model limitations and discuss possible methods for overcoming them.

### A. One-Step Approach versus Two-step Approach

The question naturally arises about the equality (or inequality) of the one-step and two-step methods. There is little experimental work on this question. Glatzel, et al., [51] compared the Mn K$\beta$ XES of $^{55}$Fe$_2$O$_3$ undergoing electron orbital capture from the 1$s$ shell with that of MnO excited by high energy x-rays. Distinct differences between the two spectra were observed, and it was proposed that multiple intermediate states need to be considered for a complete description of the process. While this suggests that the simplified process described by Eq. 3 may be insufficient for K$\beta$ XES, the approximation has not been tested extensively, and not for K$\alpha$ XES. However, if the simplified approach (only considering the lowest energy intermediate state) is indeed justified, then the greater theoretical complexity and much higher computational cost of the full excitation/de-excitation calculation can be avoided.

To address this question in detail, we compare the one-step and two-step approaches in Figure 3, where the left panel shows the energy transfer plane for NiO and PbCrO$_4$. We integrate along the diagonal of this plane ($\omega_1 - \varepsilon_{corehole} = \omega_1 - \omega_2$) to obtain the two-step prediction for the nonresonant XES. The right panel of Figure 3 compares the predictions of the one-step and two-step approaches. There is little difference between the two approaches. While the same DFT step (a few hours computation time on a modern CPU) is used for both approaches, the MLFT portion of the two-step approach requires 10 to 100 times more computational time than for the one-step approach. Given the excellent agreement between the two approaches, we use the one-step approach for all subsequent calculations.

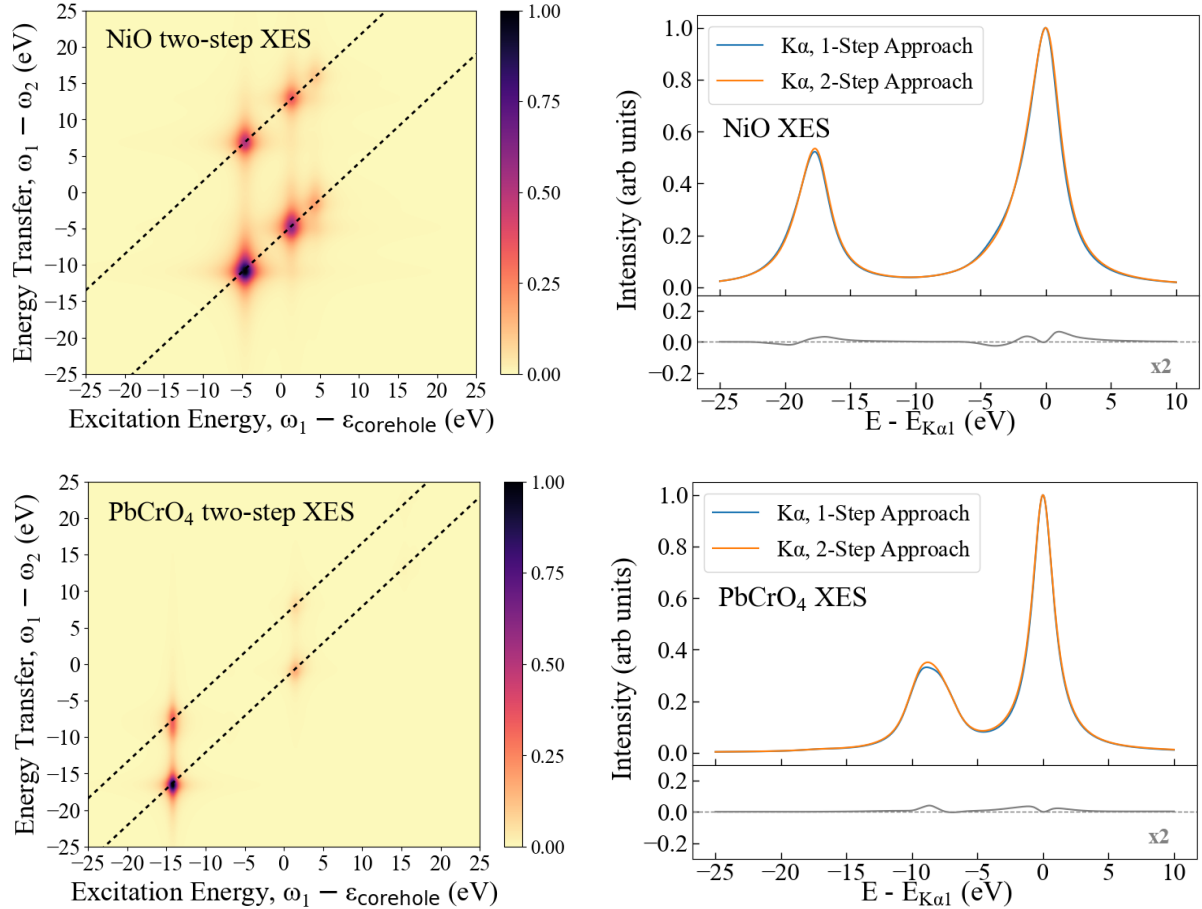

**Figure 3:** Calculated results of the one-step and two-step approaches for transition metal systems NiO and $PbCrO_4$. (Left) The energy transfer plane for both compounds. By integrating along the line corresponding to a slope of 1 (constant Emission Energy) we calculate the nonresonant XES. (Right) Calculations of the one-step approach (blue) compared to those of the two-step approach (orange), with differences between the two approaches also shown (grey).

## B. Comparison Between Theory and Experiment (Full Spectra)

Below we present a full spectral comparison between the calculated and experimental spectra for 8 different transition metal compounds. We briefly discuss the general strengths and weaknesses of the results and conclude by addressing a few interesting features of individual spectra.

| | $F2_{dd}$ | $F4_{dd}$ | $G1_{pd}$ | $G3_{pd}$ | $F2_{pd}$ | $\zeta_{2p}$ | $\zeta_{3d}$ | $\Delta$ | $U_{dd}$ | $U_{pd}$ | $U_{sd}$ | $\Gamma_{K\alpha_1}$ | $\Gamma_{K\alpha_2}$ |
|---|---|---|---|---|---|---|---|---|---|---|---|---|---|
| NiO | 11.704 | 7.230 | 5.154 | 2.929 | 6.929 | 11.510 | 0.081 | 4.700 | 7.300 | 8.500 | 8.760 | 2.060 | 2.510 |
| $CrCl_2$ | 7.856 | 4.805 | 3.019 | 1.710 | 4.318 | 5.670 | 0.050 | 6.500 | 4.500 | 5.220 | 5.400 | 1.650 | 2.090 |
| $Cr_2O_3$ | 8.531 | 5.246 | 3.216 | 1.822 | 4.585 | 5.670 | 0.050 | 5.500 | 5.500 | 6.380 | 6.600 | 1.650 | 2.090 |
| $PbCrO_4$ | 8.203 | 5.031 | 3.116 | 1.766 | 4.451 | 5.670 | 0.050 | -8.000 | 2.000 | 2.320 | 2.400 | 1.650 | 2.090 |
| $MnF_2$ | 9.575 | 5.906 | 3.877 | 2.199 | 5.388 | 6.846 | 0.053 | 6.000 | 3.000 | 3.480 | 3.600 | 1.710 | 2.320 |
| MnO | 9.649 | 5.954 | 3.898 | 2.211 | 5.415 | 6.846 | 0.053 | 8.500 | 7.000 | 8.120 | 8.400 | 1.710 | 2.320 |
| $Mn_2O_3$ | 9.658 | 5.960 | 3.900 | 2.213 | 5.419 | 6.846 | 0.053 | 6.000 | 4.000 | 4.640 | 4.800 | 1.710 | 2.320 |
| $MnO_2$ | 9.722 | 6.003 | 3.920 | 2.224 | 5.444 | 6.846 | 0.053 | 5.000 | 3.000 | 3.480 | 3.600 | 1.710 | 2.320 |

**Table 1**: The Slater-Condon, the spin-orbit splitting terms $\zeta$, charge-transfer $\Delta$ and $U_{dd}$, and lifetime broadening terms for each material.

The Slater-Condon, charge-transfer, spin-orbit coupling, and broadening parameters for each material are presented in Table 1. Of the eight materials considered, NiO and MnO are both well studied Mott insulators, and are commonly used as standard test beds for investigating highly correlated materials [52]. Both have rock-salt crystal structures and are thus highly symmetric with perfect octahedral coordination, simplifying their treatment. In contrast, $MnF_2$, $MnO_2$, $Cr_2O_3$, and $CrCl_2$ all have single site, distorted octahedral coordination. $Mn_2O_3$ is also distorted octahedral but has two unique Mn sites which must be independently considered before being averaged together. Finally, $PbCrO_4$ has a distorted tetrahedral symmetry, the lowest symmetry system being studied, in addition to being the only $3d^0$ system (nominally no $3d$ electrons).

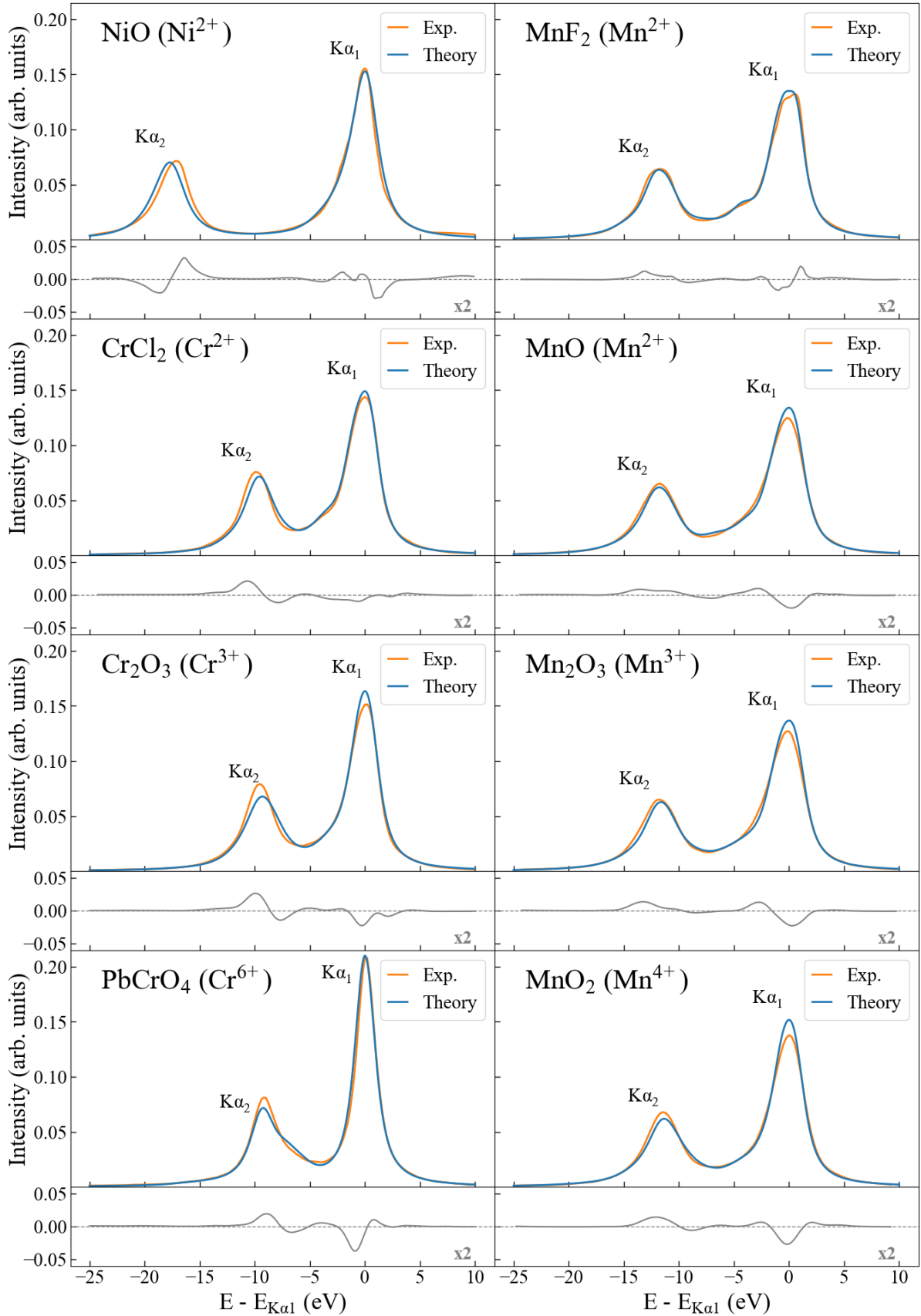

**Figure 4:** Theoretical (blue) and experimental (orange) Kα XES for several 3d transition metal compounds. Difference curves (gray) are shown in the lower panel of each subplot.

The calculated and experimental spectra for all materials are presented in Figure 4. For most of the materials studied we observe good agreement between theory and experiment, at the same level or higher compared to other DFT + MLFT calculations [29, 30, 53] and alternative *ab-initio* techniques such as CI [54] or Discrete-Variational (DV) Xα [55]. The remaining spectral discrepancies can be discussed in terms of line shape differences or small disagreements in peak-to-peak splitting.

While the overall line shapes are in good agreement, there is a systematic trend of the K$\alpha_1$ peaks being too narrow (under-broadened), the K$\alpha_2$ peaks being too wide (over-broadened), and both peaks exhibiting slightly inaccurate asymmetries. One possible explanation for these discrepancies is that the modified Slater-Condon terms are incorrect, leading to inaccurate exchange terms between the 2$p$ and 3$d$ states and thus adversely affecting the multiplet splittings. However, this would affect the K$\alpha_1$ and K$\alpha_2$ states roughly equally and therefore can be ruled out. It is tempting to attribute the broadening issue entirely to errors in the core hole lifetime values, since the overall agreement can be improved significantly by allowing these lifetimes to vary (Figure SI-V 3). However, to achieve best fits this leads to unphysical cases in which the K$\alpha_2$ broadenings are comparable to the K$\alpha_1$ broadenings, which is not possible due to Coster-Kronig decay [56]. Allowing the broadening to vary from tabulated values also causes a deviation of theory from experiment in the high energy tails of K$\alpha_1$ (Figure SI-V 4), which together with the previous point make broadening alone unlikely to be the culprit.

With these explanations being ruled out, the error in line shape is more likely due to DFT's tendency to estimate too large of a coupling between the 3$d$ and ligand states. The off-diagonal coupling terms of the $H^{TB}$ indirectly modulates the strength of the exchange interaction between the 3$d$ and 2$p$ levels by controlling the configuration of the 3$d$ shell, as well as directly modulating the energies of the valence level states. Reducing these off-diagonal coupling terms would influence multiple factors including the relative multiplet splittings and the position of any low binding energy K$\alpha_1$ satellites that sit near the K$\alpha_2$ peak. Future work will be needed to further explore this issue, with one possibility being the use of LDA + U which could help better localize the 3$d$ electrons. Finally, we note that other work has shown how more advanced MLFT models that incorporate separate Slater-Condon couplings between the various crystal field split valence states can affect the broadening of the $2p_{1/2}$ and $2p_{3/2}$ levels [21].

The peak-to-peak splitting is dominated by the relativistic $2p$ spin-orbit splitting, parameterized by the coupling term $\zeta_{2p}$, which separates emission from the $2p_{3/2}$ and $2p_{1/2}$ levels. This spin-orbit coupling is a largely intrinsic property and is not expected to change significantly when going from an atomic to solid-state system [57]. The errors in the calculated peak-to-peak splittings are on the order of a few tenths of an eV, with NiO being an outlier with a peak splitting error of ~1 eV. Although subtle changes in the Slater-Condon and charge transfer terms may explain some of the error and result in peak-to-peak splitting variations on the order of half an eV, the multiple interaction terms in the Hamiltonian make it difficult to pinpoint a single source of error. This discrepancy is also noted in empirical MLFT calculations performed by Glatzel [43].

Turning now to some intriguing features of the individual spectra, $MnF_2$ exhibits an asymmetric doublet feature in the $K\alpha_1$ as can be seen in Figure 5. This doublet splitting is approximately 1 eV and can only be seen with high resolution instruments [48]. This feature is well documented [58], and is a result of large exchange interaction allowing individual multiplet peaks to be resolved. While this same doublet feature is seen in the theory, the relative intensities of the multiplet peaks are slightly off compared to the experimental ones, causing the overall peak to broaden together into a flatter peak shape. However, this demonstrates how even in a spectroscopy with objectively simple line shapes, individual multiplet peaks can still be interpreted for the role they play [43] in much the same way as in XPS [5, 59] or RIXS [27].

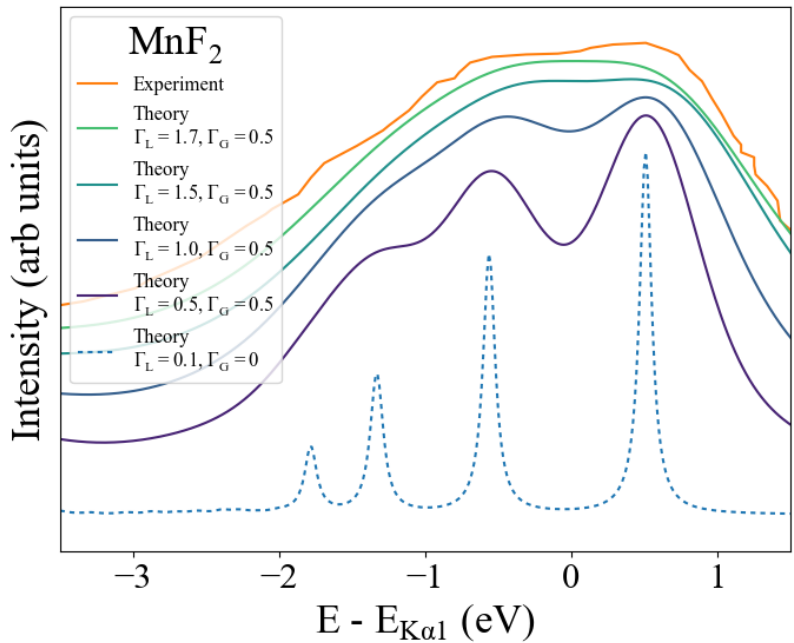

**Figure 5:** MnF$_2$ experiment versus theory for various broadenings. The multiplet doublet visible in the experimental K$\alpha_1$ peak is also present in the theoretical results, but becomes obscured by lifetime and experimental broadening.

Shifting from the details of fine spectral features to full spectral trends across compounds, to demonstrate a potential application we chose to reproduce a key result from a 2018 study by Jahrman et al. [46] which identified nonresonant XES as a promising analytical method for determining the ratio of Cr(VI) to Cr(III) in chemically challenging matrices, such as plastics. Standard wet chemical methods are susceptible to species interconversion and incomplete extraction, which has the potential to systematically underestimate the mass fraction of Cr(VI) [60, 61]. However, the K$\alpha$ XES spectra for Cr(III) and Cr(VI) are distinct enough to allow for a linear combination to be fit to an unknown composition ratio. In Figure 6 we show that we can reproduce experimentally observed differences between the Cr(III) to Cr(VI) K$\alpha$ XES using Cr$_2$O$_3$ and PbCrO$_4$, as is shown from the residuals in grey. The most notable difference is the high peaked, narrow K$\alpha_1$ signal of Cr(VI), which is a consequence of the 6+ formal oxidation state, leading to a distinct difference in the FWHM and intensities of the K$\alpha_1$ peaks.

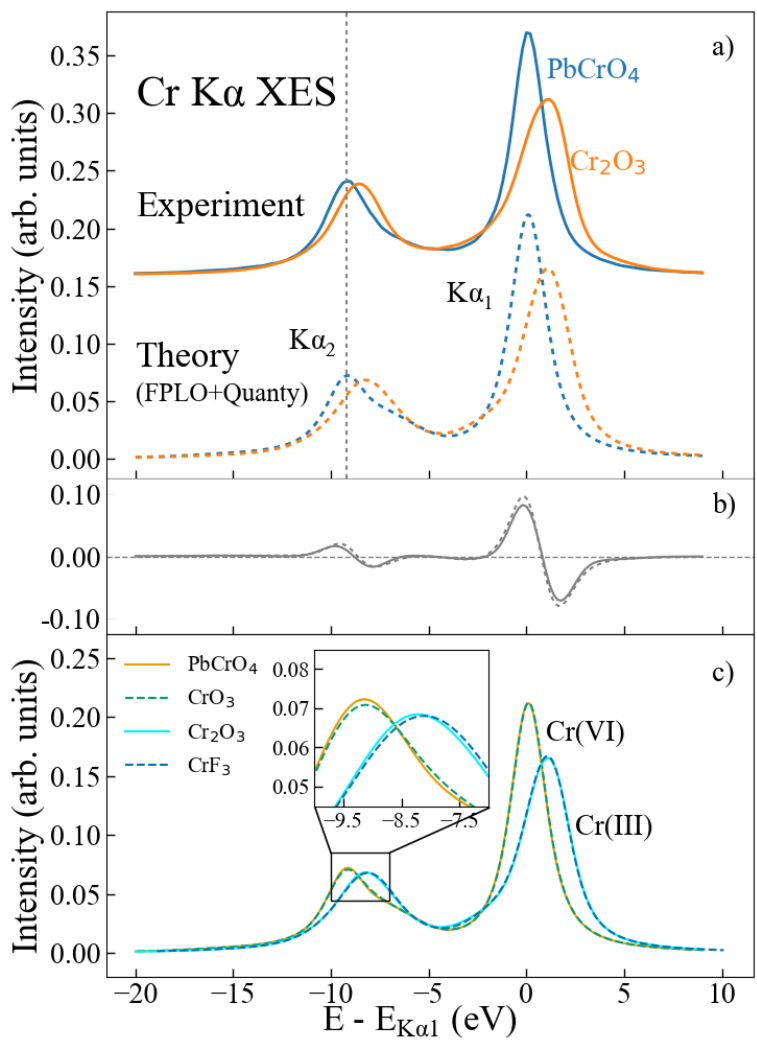

**Figure 6:** (a) Experimental (solid) and theory (dashed) spectra for $Cr_2O_3$ and $PbCrO_4$.. (b) Difference curves for Cr(VI) experiment – Cr(III) experiment (solid) and Cr(VI) theory – Cr(III) theory (dashed). (c) Theoretical XES spectra for two Cr(III) compounds ($CrF_3$ and $Cr_2O_3$) and two Cr(VI) compounds ($Cr_2O_3$ and $PbCrO_4$).

This distinction between $Cr_2O_3$ and $PbCrO_4$ Kα XES was demonstrated by Jahrman et al. [46] to be constant across Cr(III) and Cr(VI) materials. We found that this observation also held true for our calculated spectra, as demonstrated in the bottom of Figure 6. Both of the Cr(VI) compounds are tetragonally coordinated with oxygen and the Cr(III) compounds are octahedrally coordinated with fluorine and oxygen respectively. We note that the simulated Kα XES spectra are relatively insensitive to the particular compound, and are instead sensitive to the nominal oxidation state and environmental symmetry, again agreeing with trend observed in experiment by Jahrman et al. [46].

## C. Comparison Between Experiment and Theory (Spectral Characteristics)

Neglecting interactions, the Kα$_1$:Kα$_2$ integral ratio is naively expected to be 2:1 due to the 4:2 ratio of occupancy of the $2p_{3/2}$ and $2p_{1/2}$ shells. While this is a reasonable starting assumption, the ratio is skewed by mixing between these states due to Coulomb interactions between the $2p$ core hole and $3d$ electrons[59]. In the top panel of Figure 7 we compare theory and experiment. The deviation from the 2:1 ratio is most clear for CrCl$_2$, MnO and Mn$_2$O$_3$, each of which have strong shoulders on the lower energy side of Kα$_1$. We note that the extracted ratio is highly dependent on the quality of the Voigt fits that are used to deconvolve the Kα$_1$ and Kα$_2$ peaks (Figure SI-IV 1). The scatter plot in Figure 8 (a) shows the extent of correlation between theory and experiment.

The FWHM of the peaks are a common metric for characterizing the Kα XES [43, 47, 49], even though the presence of many distributed multiplet features can impede its easy interpretation. As with the Kα$_1$:Kα$_2$ ratio there is generally good correlation between theory and experiment; see also the bottom panel of Figure 7 and Figure 8 (b). The strongest outlier is the PbCrO$_4$ Kα$_2$ FWHM. This, as well PbCrO$_4$'s low Kα$_1$:Kα$_2$ ratio, can be attributed to a combination of the $2p$, $3d$ Coulomb exchange interaction and valence level hybridization. These issues will be discussed in greater detail in section IV.D, where we address limitations of the model.

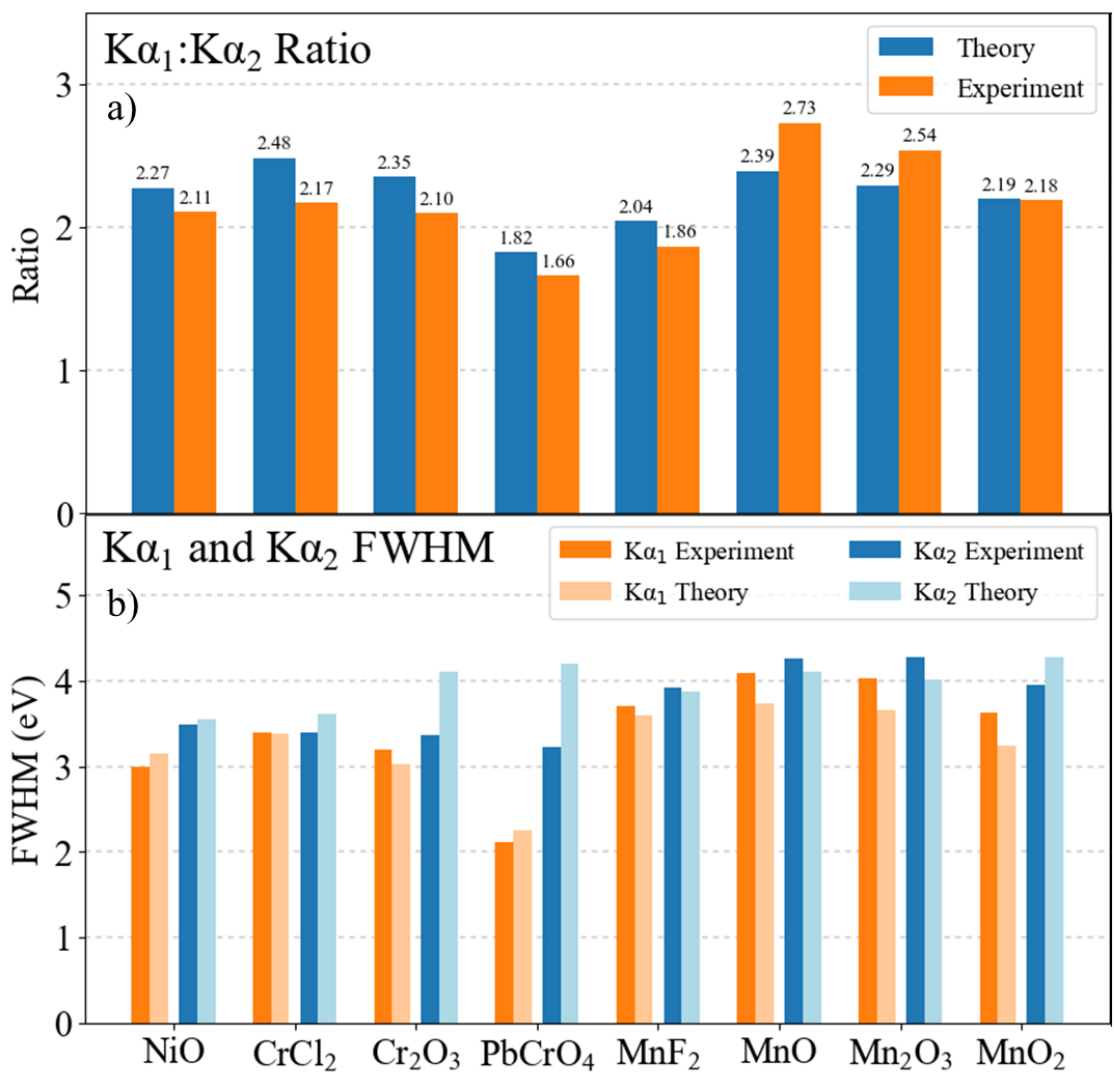

**Figure 7:** Comparison of theory and experiment for (a) $K\alpha_1$:$K\alpha_2$ integral ratio and (b) FWHM for the various compounds. The FWHM values were calculated according to the procedure in Figure SI-V 1.

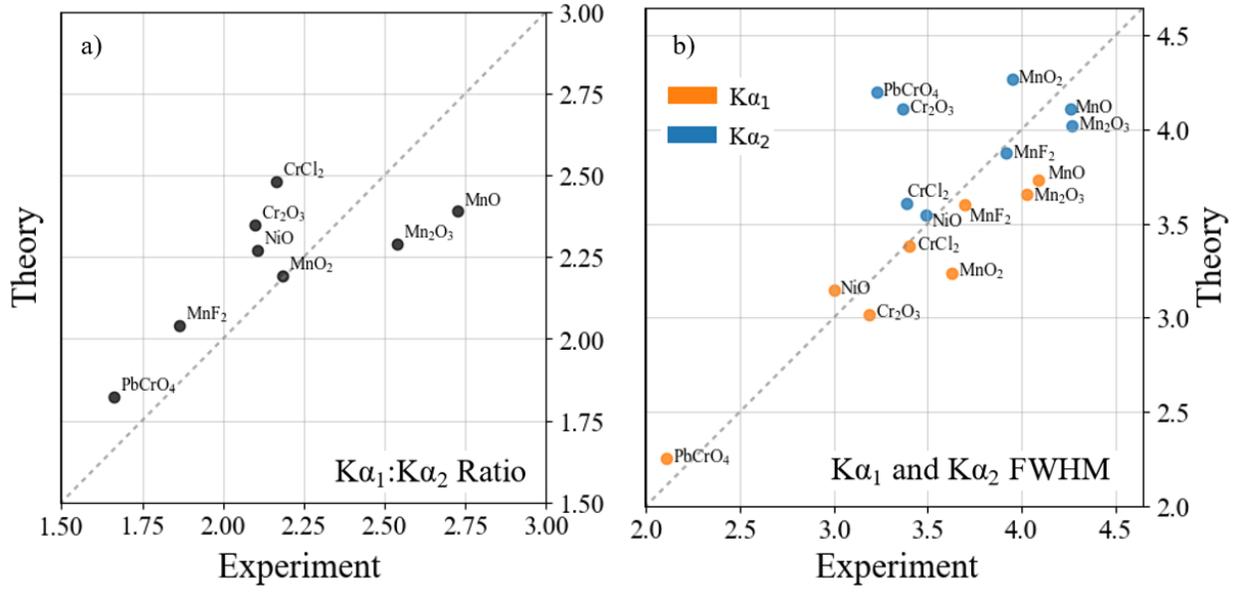

**Figure 8:** Trendlines between theory and experiment for the Kα$_1$:Kα$_2$ integral ratio and FWHM.

Using the FWHM values extracted from the calculated spectra, we can now explore the often-stated rule of multiplet theory that Kα$_1$ linewidth is directly proportional to the exchange interaction between the 2p and 3d orbitals, $G^1_{pd}$ and $G^3_{pd}$ [6]. This will allow us to test a known experimental trend across a set of compounds and confirm an important caveat first raised by Kawai [44]. Specifically, the charge-transfer state in the presence of the core hole will necessarily influence the Kα$_1$ FWHM because the number of unpaired 3d electrons will change in response to the core hole potential. Determining whether the theory follows the established FWHM trend provides an important litmus test, as the same charge-transfer parameters $U_{dd}$ and Δ which determine the fits to experiment also determine the number of unpaired 3d electrons in the intermediate state.

This dependence on the core hole $n_{unpaired\ 3d}$ is the origin of the similarity between the Kα XES of compounds such as MnO and Mn$_2$O$_3$ [43], and FeO and Fe$_2$O$_3$ [44], with their nearly identical metal ion spin states even though they have different classical oxidation. The derivation for the linear relationship between the Kα$_1$ FWHM and the number of unpaired *3d* electrons relies on the use of a free ion model [62] and has a number of key assumptions, most notably that the orbital angular momentum of the *3d* states is quenched ($J = S$), simplifying the coupling between the final state *2p* hole and the valence shell [43]. As such, this linear trend is expected to be more of a general rule, rather than a hard and fast law.

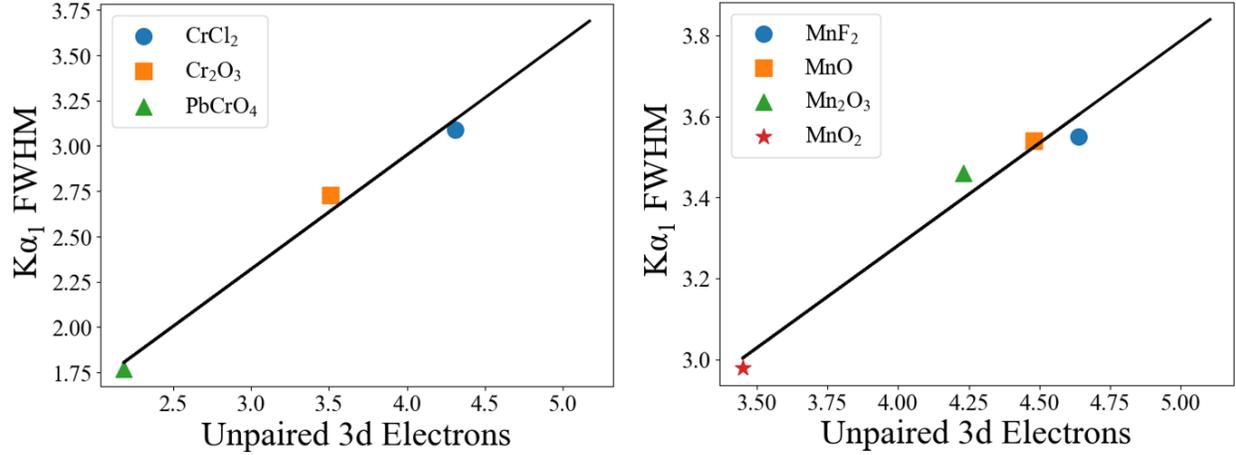

**Figure 9:** Kα$_1$ FWHM vs Number of Unpaired 3d Electrons. Reported values of FWHM are calculated from theorical results broadened with lifetime values reported in Table 1. Both the Cr (left) and Mn (right) compounds exhibit the expected linear trend.

| | Nominal # of 3d Electrons | $\langle n_d \rangle$ | $\langle S^2 \rangle$ | # of Unpaired 3d Electrons | Kα$_1$ FWHM (only lifetime broadening) |
|---|---|---|---|---|---|
| NiO | 8.00 | 9.03 | 0.79 | 1.04 | 2.76 |
| CrCl$_2$ | 4.00 | 4.88 | 6.81 | 4.31 | 3.09 |
| Cr$_2$O$_3$ | 3.00 | 4.35 | 4.83 | 3.51 | 2.73 |
| PbCrO$_4$ | 0.00 | 5.43 | 2.28 | 2.18 | 1.77 |
| MnF$_2$ | 5.00 | 5.38 | 7.71 | 4.64 | 3.55 |
| MnO | 5.00 | 5.50 | 7.25 | 4.48 | 3.54 |
| Mn$_2$O$_3$ | 4.00 | 4.90 | 6.59 | 4.23 | 3.46 |
| MnO$_2$ | 3.00 | 4.68 | 5.18 | 3.66 | 2.98 |

**Table 2:** Expectation values for the $n_d$ number operator and $S^2$ operator, both calculated over the 3d fermionic modes. The number of unpaired 3d electrons (calculated from Eq. 6) in the presence of a core hole, the Kα$_1$ FWHM (calculated from theory), and the nominal number of total 3d electrons are also reported.

The number of unpaired 3$d$ electrons in the presence of the core hole and the Kα$_1$ FWHM values are reported in Table 2 and are plotted in Figure 9 for the Cr and Mn series. As addressed earlier, PbCrO$_4$ is nominally a d$^0$ system. However, when hybridization with the ligand orbitals is considered, the ground state occupation's expectation value becomes approximately 1, which increases to approximately 2 when considering the effect of charge-transfer in the presence of the core hole. The expected linear relationship is only observed when considering these multiconfigurational effects, confirming Kawai's findings that they are necessary for accurately representing the excited magnetic state of the system [44].

### D. Effect of $\Delta$ and $U_{dd}$ on the spectrum, and limitations of MLFT

As noted above a long-term goal of this research program is a fully *ab-initio*, predictive treatment of Kα XES with no adjustable parameters. Here, the traditional MLFT parameter space has been reduced to only two dimensions; however, the spectra are still quite sensitive to $\Delta$ and $U_{dd}$. The interplay of the tuning parameters for charge transfer and Coulomb interactions adds complexity to the spectra, making it difficult to isolate the origin of certain spectral features. However, the reduced parameter space still simplifies the process of fitting theory to experiment and allows more detailed exploration than previously would be possible.

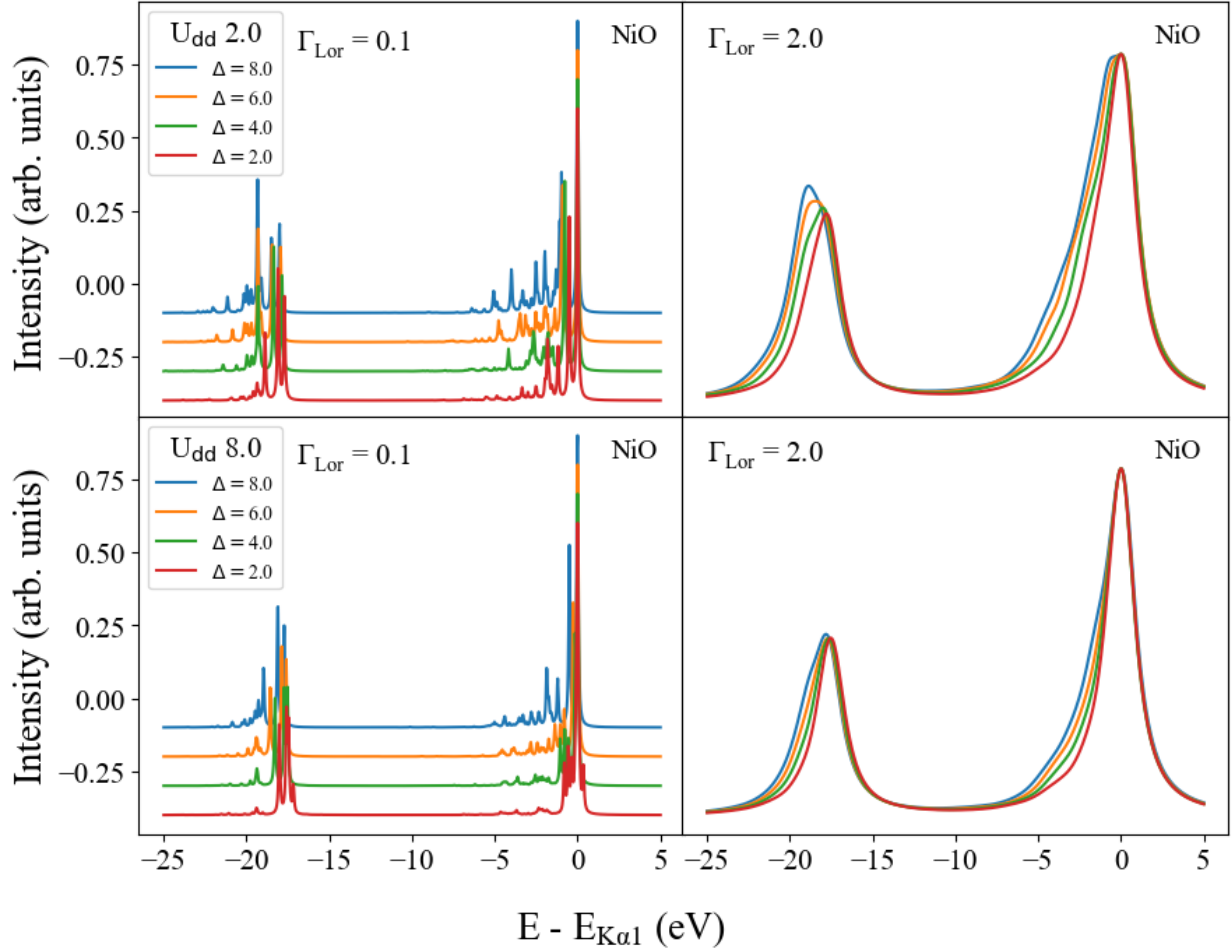

**Figure 10:** Parameter space exploration of NiO. The left panels have Lorentzian broadening of only $\Gamma_{Lor} = 0.1$ eV while the right panels have $\Gamma_{Lor} = 2.0$ eV.

In Figure 10 we investigate the effect of $U_{dd}$ and $\Delta$ on DFT + MLFT calculations for NiO. Scaling $\Delta$ has the effect of widening the distance between the *d* and ligand centroids, thereby increasing the energy of the gap between occupied and unoccupied valence states. This is the origin of the increase in multiplet peak splitting with $\Delta$ as seen in Figure 10 and results in low-energy tails in the broadened spectra in the right column. Ideally these can be dealt with by fitting to the features which do appear in experimental spectra (additional shoulders or asymmetry), but this is not a trivial task as even within this reduced parameter space different unique choices of $U_{dd}$ and $\Delta$ can still produce similar spectra, making the correct parameter choice ambiguous.

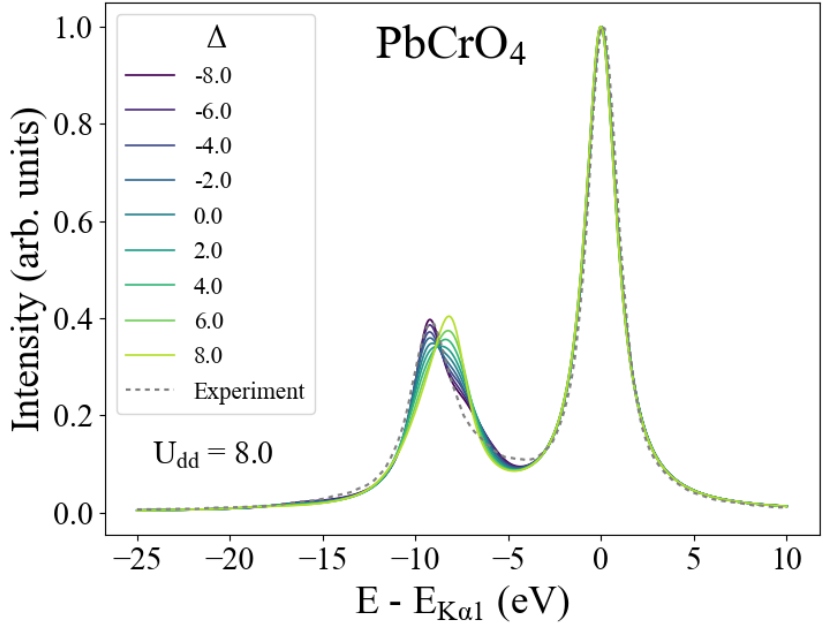

**Figure 11:** PbCrO$_4$ spectra for $U_{dd} = 8$ and $-8 \leq \Delta \leq 8$. The experimental spectrum is shown as a grey dashed line for reference.

    The extremes of the DFT + MLFT model are well demonstrated by PbCrO$_4$. The Cr in PbCrO$_4$ has nominally zero $d$ electrons, with $20!/(10!)^2 \sim 180,000$ possible electron configurations, making this the most computationally expensive calculation of all the materials in this manuscript. By scanning over $\Delta$, as shown in Figure 11, it becomes clear that the double peak feature underneath K$\alpha_2$ is modulated by the charge-transfer energy while the K$\alpha_1$ is entirely insensitive to changes in $\Delta$. In non-DFT augmented MLFT, negative values of $\Delta$ are uncommon but do occur for highly oxidized systems with more covalent character [63]. The lack of sensitivity of K$\alpha_1$ to $\Delta$ is surprising, given that number of unpaired $3d$ electrons is also modulated by $\Delta$. As $\Delta$ increases (and $U_{dd}$ is held constant), the splitting between the $d$ and ligand levels increases and the lowest energy valence states have less $d$ character, leading to a decrease in both the total number of $3d$ electrons and the number of unpaired $3d$ electrons in the intermediate ground state. This would appear to contradict the earlier conclusion about the relationship between the number of unpaired $3d$ electrons and the K$\alpha_1$ FWHM. However, the best fit between experiment and theory for PbCrO$_4$ (which depended entirely on variability in K$\alpha_2$) also resulted in a $\Delta$ value which gave the appropriate number of unpaired $3d$ electrons.

# Conclusion

We have demonstrated that a DFT + MLFT approach based on the Quanty + FPLO framework can achieve a nearly first principles calculation of core-to-core XES of 3$d$ TM systems, with only a few undetermined coefficients. In so doing, we have successfully reproduced qualitative and quantitative trends in Kα spectra with regards to oxidation and spin state. In every metric explored, theory trended strongly to agree with experiment, demonstrating that a large subset of traditionally empirical parameters can be accurately reproduced by DFT.

Many extensions of this work are possible. For example, future work could consider constrained DFT calculations which better account for the presence of the 1s core hole on the intermediate state. This may be accomplished by removing a 1s electron from an impurity transition metal atom in the DFT calculation to better approximate the radial wave functions of the valence orbitals during the intermediate state. Additionally, constrained DFT may offer another avenue for calculating both $\Delta$ and $U_{dd}$ [64]. Further developments of DFT + DMFT techniques [31], may allow for a better representation of the hybridization between the ligand and transition metal ions. In particular, future work should have a strong focus on accurate *ab-initio* methods for calculating charge-transfer parameters $U_{dd}$ and $\Delta$, as well as extending the DFT + MLFT approach to other highly correlated systems such as 4$d$ transition metals, lanthanides, and actinides.

# Acknowledgements

We thank Yiming Chen, Paul Bagus, Marius Retegan, and Maurits Haverkort for fruitful discussions about the appropriate methods for calculating XES. We would also like to thank Simon Heinze and Martin Bras for the many hours spent helping debug and explain Quanty scripts. JJK and JJR acknowledge support by DOE BSE Contract DE-AC02-76SF00515. CAC was supported by the National Science Foundation Graduate Research Fellowship Program under Grant No. DGE-2140004. Any opinions, findings, and conclusions or recommendations expressed in this material are those of the author(s) and do not necessarily reflect the views of the National Science Foundation. GTS acknowledges support from the U.S. Department of Energy, Nuclear Energy University Program under contract DE-NE0009158.

# Citations

20. Haverkort, M.W., M. Zwierzycki, and O.K. Andersen, *Multiplet ligand-field theory using Wannier orbitals.* Physical Review B, 2012. **85**(16): p. 165113.
21. Krüger, P. *First-Principles Calculation of Ligand Field Parameters for L-Edge Spectra of Transition Metal Sites of Arbitrary Symmetry*. Symmetry, 2023. **15**, DOI: 10.3390/sym15020472.
22. Ikeno, H., et al., *Multiplet calculations of L2,3 x-ray absorption near-edge structures for 3d transition-metal compounds.* Journal of Physics: Condensed Matter, 2009. **21**(10): p. 104208.
23. Abu-Samak, M., et al., *Electronic structure and energy gaps evaluation of perovskite manganite single crystals using XES and XAS spectroscopy.* Journal of Electron Spectroscopy and Related Phenomena, 2021. **250**: p. 147084.
24. Lafuerza, S., et al., *Chemical Sensitivity of Kβ and Kα X-ray Emission from a Systematic Investigation of Iron Compounds.* Inorganic Chemistry, 2020. **59**(17): p. 12518-12535.
25. Fazinić, S., et al., *Chemical sensitivity of the Kα X-ray emission of Ti and Cr compounds induced by 2 MeV protons.* Spectrochimica Acta Part B: Atomic Spectroscopy, 2022. **195**: p. 106506.
26. Miedema, P.S., et al., *Iron 1s X-ray photoemission of Fe2O3.* Journal of Electron Spectroscopy and Related Phenomena, 2015. **203**: p. 8-13.
27. Zimmermann, P., M.O.J.Y. Hunault, and F.M.F. de Groot, *1s2p RIXS Calculations for 3d Transition Metal Ions in Octahedral Symmetry.* Journal of Spectroscopy, 2018. **2018**.
28. Zimmermann, P., et al., *Quanty4RIXS: a program for crystal field multiplet calculations of RIXS and RIXS–MCD spectra usingQuanty.* Journal of Synchrotron Radiation, 2018. **25**(3): p. 899-905.
29. Md. Nur Hasan, F.S., Nivedita Pan, Dibya Phuyal, Mahmoud Abdel-Hafiez, Samir Kumar Pal, Anna Delin, Patrik Thunström, D. D. Sarma, Olle Eriksson, Debjani Karmakar, *Re-Dichalcogenides: Resolving Conflicts of Their Structure–Property Relationship.* Advanced Physics Research, 2022.
30. Gorelov, E., et al., *MLFT approach with p-d hybridization for ab initio simulations of the pre-edge XANES.* Radiation Physics and Chemistry, 2020. **175**.
31. Lüder, J., et al., *Theory of L-edge spectroscopy of strongly correlated systems.* Physical Review B, 2017. **96**(24).
32. Agrestini, S., et al., *Long-range interactions in the effective low-energy Hamiltonian of Sr2IrO4: A core-to-core resonant inelastic x-ray scattering study.* Physical Review B, 2017. **95**(20): p. 205123.
33. Holden, W.M., G.T. Seidler, and S. Cheah, *Sulfur Speciation in Biochars by Very High Resolution Benchtop Kα X-ray Emission Spectroscopy.* The Journal of Physical Chemistry A, 2018. **122**(23): p. 5153-5161.
34. Haverkort, M. *Quanty - a quantum many body scripting language*. 2022; Available from: https://quanty.org/.
35. Koepernik, K. and H. Eschrig, *Full-potential nonorthogonal local-orbital minimum-basis band-structure scheme, version 14.00-49 [https://www.FPLO.de]*. Physical Review B, 1999. **59**(3): p. 1743-1757.
36. Marguí, E., I. Queralt, and E. de Almeida, *X-ray fluorescence spectrometry for environmental analysis: Basic principles, instrumentation, applications and recent trends.* Chemosphere, 2022. **303**: p. 135006.
37. Kramers, H.A. and W. Heisenberg, *Über die Streuung von Strahlung durch Atome.* Zeitschrift für Physik, 1925. **31**(1): p. 681-708.
38. Mortensen, D.R., et al., *Benchmark results and theoretical treatments for valence-to-core x-ray emission spectroscopy in transition metal compounds.* Physical Review B, 2017. **96**(12).
39. Roychoudhury, S., et al., *Changes in polarization dictate necessary approximations for modeling electronic deexcitation intensity: Application to x-ray emission.* Physical Review B, 2022. **106**(7).
40. Heinze, S., *Material Specific Simulations of Many-Body Electron Dynamics*, in *Combined Faculty of Natural Sciences and Mathematics*. 2021, Heidelberg University: Heidelberg.
27

# Core-to-Core X-ray Emission Spectra from Wannier Based Multiplet Ligand Field Theory (Supplemental Information)

Charles A. Cardot, Joshua J. Kas, Jared E. Abramson, John J. Rehr and Gerald T. Seidler

Physics Department, University of Washington, Seattle WA

## I. Quanty and FPLO Input Files

The input required for running the CrCl$_2$ Kα XES example can be found at https://github.com/Seidler-Lab/CrCl2KaXES_Example. This contains three input files that when combined create the entire pipeline for the calculation of CrCl$_2$. All of these are modified from tutorials taken from the Quanty website [34].

## II. Green's Function Formalism

The formalism motivated here can be readily used to test the relevance of multiple intermediate states. Eq. 1 in the manuscript is often expressed in terms of an effective one particle Green's functions [65], and we will do so here as well. The two-step approach relies on the use of a third order Green's function,

$$\mathcal{G}(\omega_1, \omega_2) = \langle i| \hat{s}^\dagger \frac{1}{\omega_1 - i\frac{\Gamma_s}{2} - H_1} \hat{p}^\dagger \hat{s} \frac{1}{\omega_1 - \omega_2 - i\frac{\Gamma_p}{2} - H_2} \hat{s}^\dagger \hat{p} \frac{1}{\omega_1 - i\frac{\Gamma_s}{2} - H_1} \hat{s} |i\rangle. \quad \text{(SII Eq 1)}$$

where for Kα XES the $\hat{s}$ annihilation operator acting on the $1s$ state (initial → intermediate) and $T_{sp}$ is the dipole transition operator between the $2p$ and $1s$ states (intermediate → final). We do not include the $\vec{\epsilon} \cdot \vec{r}$ term with the understanding that all polarization directions must be considered for an isotropic spectrum. In SII Eq. 1, $H_1$ and $H_2$ correspond to the Hamiltonians of the systems with a $1s$ core hole and $2p$ core hole respectively, with $i$ being the set of ground state wavefunctions which come from solving the eigenstates of the initial state Hamiltonian. The response function has arguments $\omega_1$ and $\omega_2$ which are defined the same way that they are in Eq. 1, corresponding to incident energy and emission energy, respectively. Making the same approximation that was made when going from Eq. 1 to Eq. 2, SII Eq 1. can be rewritten using a one-step approach with a first order Green's function,

$$\mathcal{G}(\omega_2) = \langle i'|\hat{p}^\dagger \hat{s} \frac{1}{\omega_2 - i\frac{\Gamma}{2} - H} \hat{s}^\dagger \hat{p}|i'\rangle, \quad \text{(SII Eq 2)}$$

where now the $i'$ states describe the intermediate ground state in the presence of the 1s core hole.

## III. Ligand Field Equations

Ligand field equations for the initial, intermediate, and final states of the Kα XES process, where $\varepsilon_s, \varepsilon_p, \varepsilon_L$, and $\varepsilon_d$ are the average onsite energies of the $s$, $p$, Ligand, and $d$ orbitals respectively. Solving the ligand field equations for the initial, intermediate, and final states allows us to define the onsite energy values in terms of $\Delta$ and $U_{dd}$, which are then used to constrain the intermediate and final state Hamiltonians respectively.

**Initial State:**

$$10\varepsilon_L + n\varepsilon_d + n(n-1)\frac{U_{dd}}{2} = 0$$

$$9\varepsilon_L + (n+1)\varepsilon_d + (n+1)n\frac{U_{dd}}{2} = \Delta$$

$$8\varepsilon_L + (n+2)\varepsilon_d + (n+2)(n+1)\frac{U_{dd}}{2} = 2\Delta + U_{dd}$$

**Intermediate State:**

$$2\varepsilon_s + 10\varepsilon_L + n\varepsilon_d + n(n-1)\frac{U_{dd}}{2} + 2nU_{sd} = 0$$

$$2\varepsilon_s + 9\varepsilon_L + (n+1)\varepsilon_d + (n+1)n\frac{U_{dd}}{2} + 2(n+1)U_{sd} = \Delta$$

$$2\varepsilon_s + 8\varepsilon_L + (n+2)\varepsilon_d + (n+2)(n+1)\frac{U_{dd}}{2} + 2(n+2)U_{sd} = 2\Delta + U_{dd}$$

$$\varepsilon_s + 10\varepsilon_L + n\varepsilon_d + n(n-1)\frac{U_{dd}}{2} + nU_{sd} = 0$$

$$\varepsilon_s + 9\varepsilon_L + (n+1)\varepsilon_d + (n+1)n\frac{U_{dd}}{2} + (n+1)U_{sd} = \Delta - U_{sd}$$

$$\varepsilon_s + 8\varepsilon_L + (n+2)\varepsilon_d + (n+2)(n+1)\frac{U_{dd}}{2} + (n+2)U_{sd} = 2\Delta + U_{dd} - 2U_{sd}$$

**Final State:**

$$6\varepsilon_p + 10\varepsilon_L + n\varepsilon_d + n(n-1)\frac{U_{dd}}{2} + 6nU_{pd} = 0$$

$$6\varepsilon_p + 9\varepsilon_L + (n+1)\varepsilon_d + (n+1)n\frac{U_{dd}}{2} + 6(n+1)U_{pd} = \Delta$$

$$6\varepsilon_p + 8\varepsilon_L + (n+2)\varepsilon_d + (n+2)(n+1)\frac{U_{dd}}{2} + 6(n+2)U_{pd} = 2\Delta + U_{dd}$$

$$5\varepsilon_p + 10\varepsilon_L + n\varepsilon_d + n(n-1)\frac{U_{dd}}{2} + 5nU_{pd} = 0$$

$$5\varepsilon_p + 9\varepsilon_L + (n+1)\varepsilon_d + (n+1)n\frac{U_{dd}}{2} + 5(n+1)U_{pd} = \Delta - U_{pd}$$

$$5\varepsilon_p + 8\varepsilon_L + (n+2)\varepsilon_d + (n+2)(n+1)\frac{U_{dd}}{2} + 5(n+2)U_{pd} = 2\Delta + U_{dd} - 2U_{pd}$$

# IV. Voigt Fits

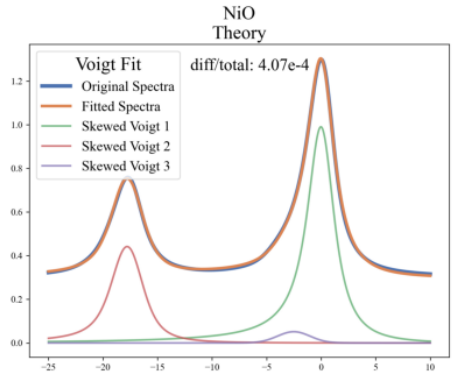
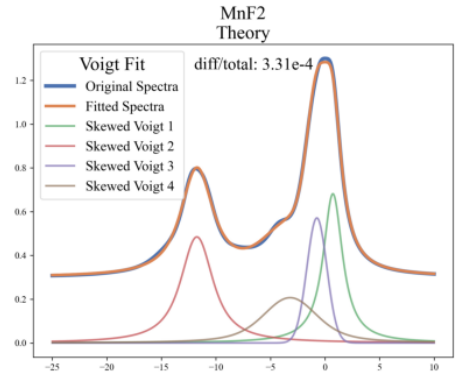
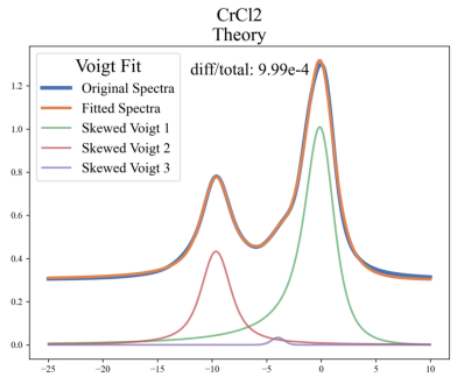
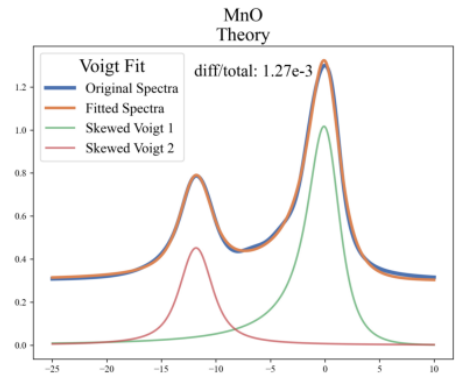
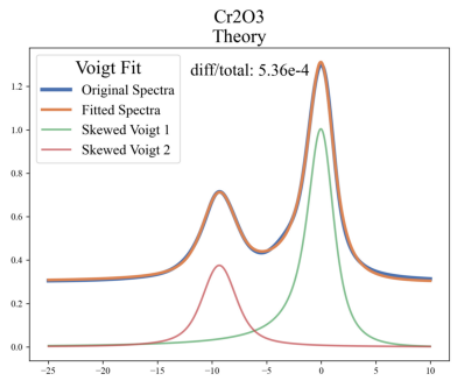
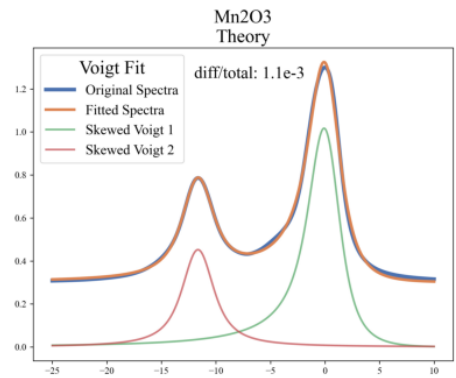
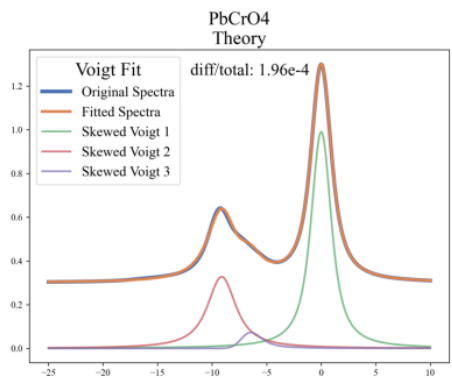
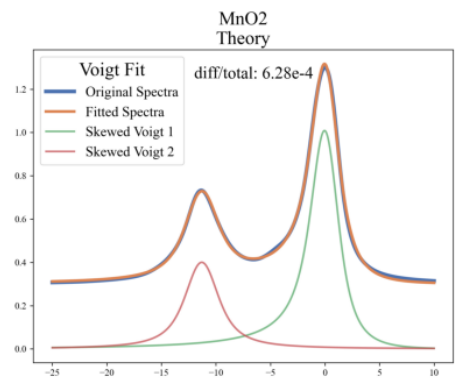

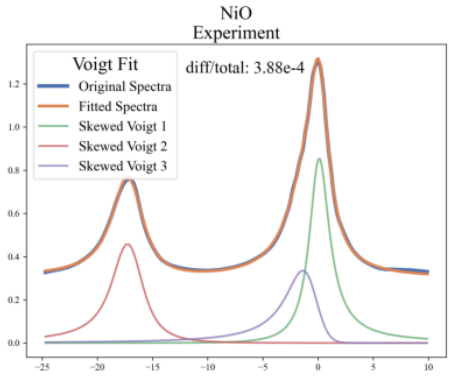
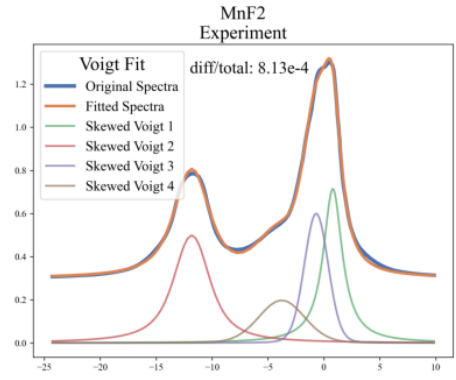
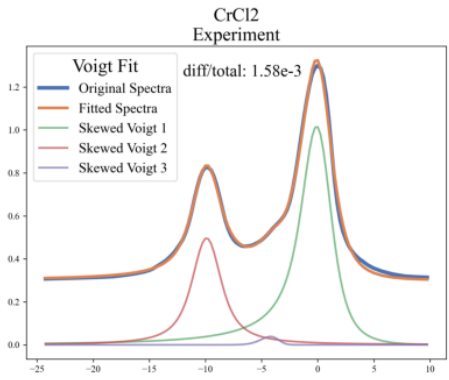
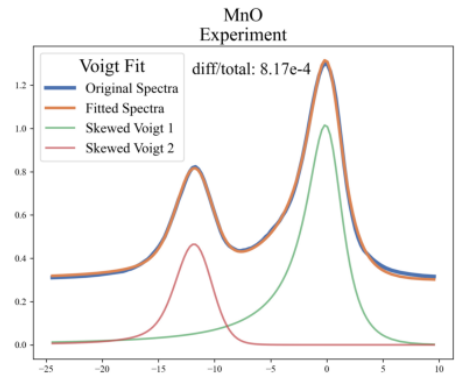
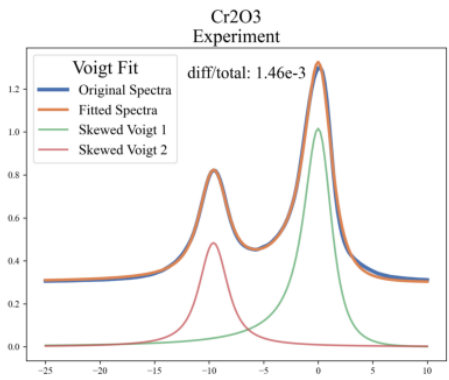
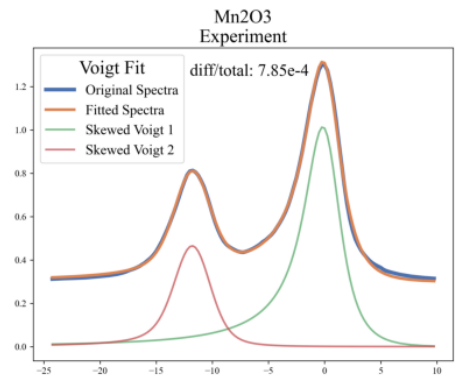
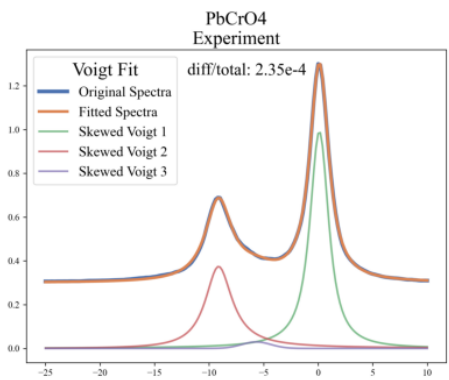
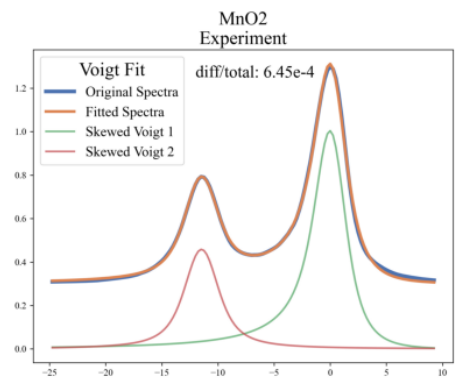

**Figure SI-IV 1:** Skewed Voigt fits comprised of 2, 3, or 4 individual peaks for both calculated and experimental spectra. The number of individual functions used for a particular material were the same for calculated and experiment. The diff/total parameter relates to the integrated absolute difference divided by the total area under the curve and gives some measure of how "good" the Voigt fit is.

## V. FWHM Calculation and Broadening Schemes

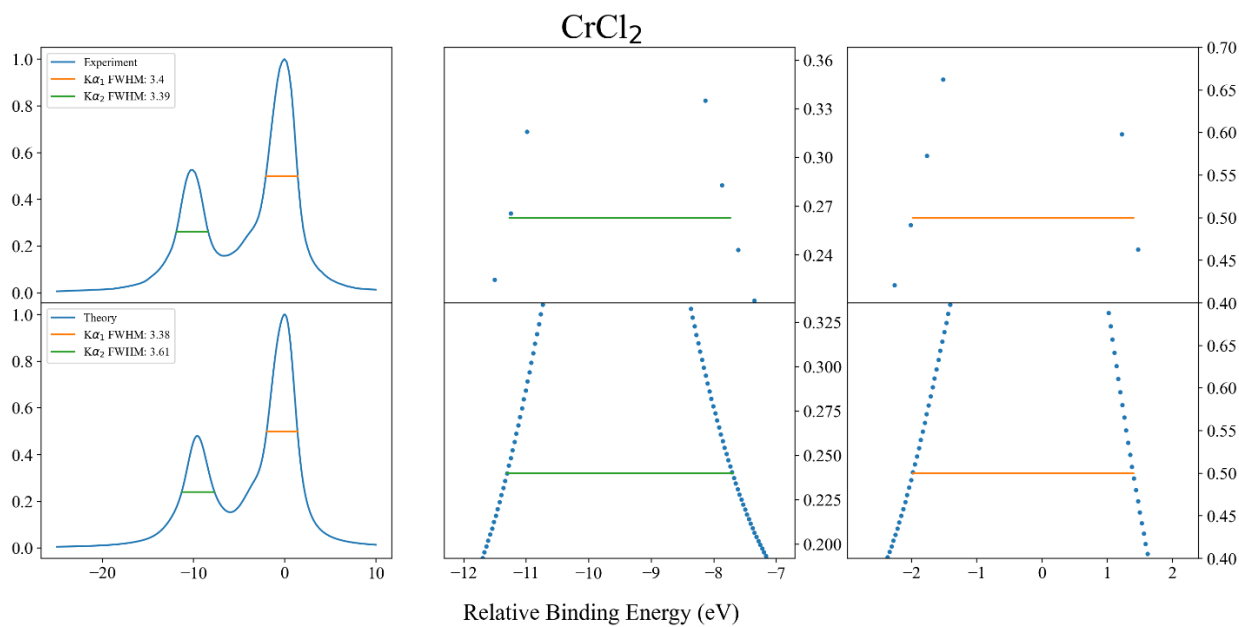

**Figure SI-V 1:** Depiction of how the FWHM values were calculated for the Kα$_1$ and Kα$_2$ peaks. The full width at half maximum was calculated via linear interpolation between data points.

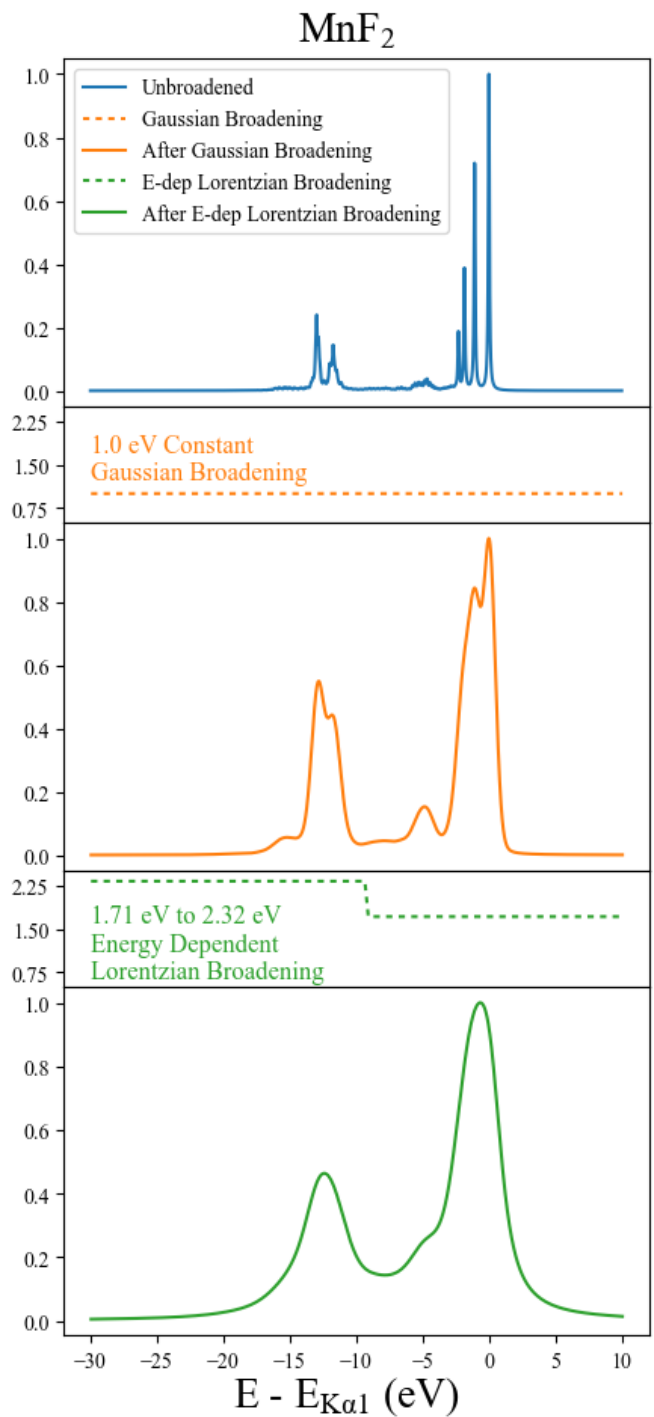

**Figure SI-V 2:** Depiction of the how the experimental and lifetime broadenings were applied for $MnF_2$. Due to the commutative nature of convolutions, the order in which they are applied does not matter.

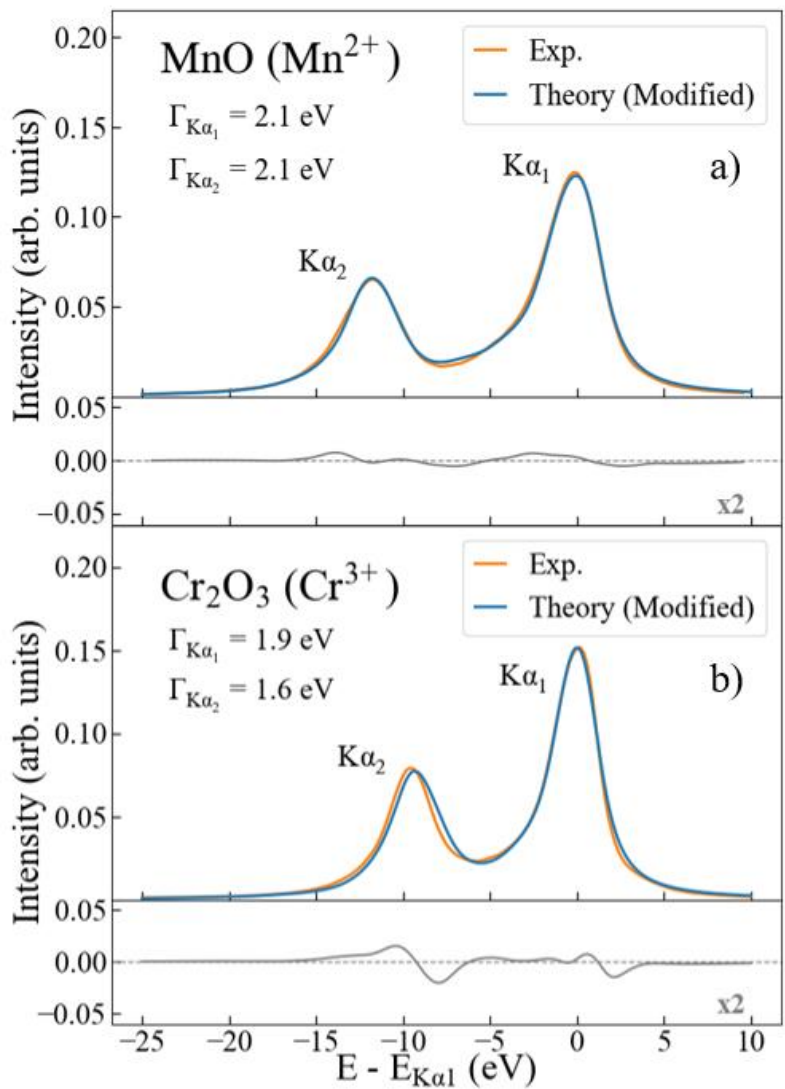

**Figure SI-V 3:** Modified theoretical and experimental Kα for (a) MnO and (b) $Cr_2O_3$. The theoretical spectra have been modified by changing the lifetime broadening values for better overall agreement between theory and experiment. To original lifetime broadening values can be found in Table 1.

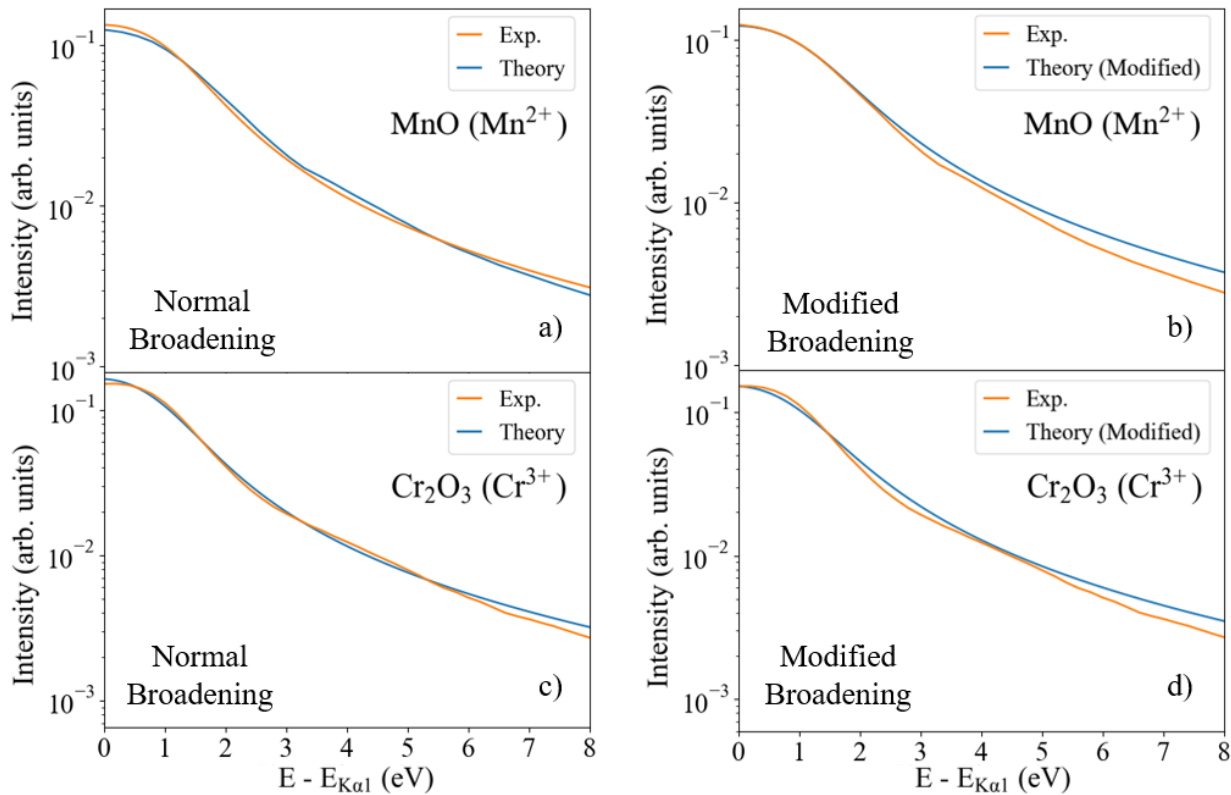

**Figure SI-V 4:** A zoomed in comparison of the theory and experiment for MnO (a, b) and $Cr_2O_3$ (c, d) with normal (a, c) and modified (b, d) broadening. The plots with modified broadening show that the larger lifetime broadenings on the $K\alpha_1$ high binding energy side diverge from the experimental line shape for both compounds.